\documentclass[pdflatex,sn-mathphys-num]{sn-jnl}

\usepackage{graphicx}%
\graphicspath{{figures/main/}{figures/extended/}}
\usepackage{multirow}%
\usepackage{amsmath,amssymb,amsfonts}%
\usepackage{amsthm}%
\usepackage{mathrsfs}%
\usepackage[title]{appendix}%
\usepackage{xcolor}%
\usepackage{textcomp}%
\usepackage{manyfoot}%
\usepackage{booktabs}%
\usepackage{algorithm}%
\usepackage{algorithmicx}%
\usepackage{algpseudocode}%
\usepackage{listings}%
\usepackage{float}
\usepackage{siunitx}
\usepackage{afterpage}

\begin{document}

\title[Article Title]{Accelerated descriptor-free path sampling for protein--ligand binding kinetics}

\author[1,2]{\fnm{Simon M.} \sur{Lichtinger}}
\author*[1,2]{\fnm{Roberto} \sur{Covino}}\email{covino@fias.uni-frankfurt.de}

\affil[1]{\orgdiv{Institute for Computer Science}, \orgname{Goethe University Frankfurt}, \country{Germany}}
\affil[2]{\orgdiv{Frankfurt Institute for Advanced Studies (FIAS)}, \orgname{Goethe University Frankfurt}, \country{Germany}}

\abstract{The kinetics of protein--ligand binding systems are increasingly recognized as a key determinant of drug efficacy, yet remain far harder to compute than binding affinities. Existing kinetics methods either bias the dynamics along a collective variable (CV), demanding careful system-specific CV design, or use path sampling, which keeps the dynamics unbiased but can struggle to converge rates out of deep free-energy wells and often relies on hand-engineered descriptors. By combining the `best of both worlds', we propose a method to compute accurate kinetics for general ligand-unbinding problems at modest computational expense and minimal fine tuning, building on the AI for Molecular Mechanism Discovery (AIMMD) path sampling framework. To avoid the need for feature engineering, we opt for modelling the committor with a single descriptor-free, equivariant graph neural network shared across all systems. We also partially flatten deep bound-state wells with a static, basin-restricted bias potential. This improves convergence by lifting the path sampling state boundary out of regions, where the committor is hard to learn, while leaving the reactive region strictly unbiased. Across host--guest and protein--ligand systems spanning roughly 17 orders of magnitude in residence time, the method robustly recovers rates in line with reference and experimental values. Simultaneously, and without further sampling, it also reconstructs the underlying unbinding mechanisms. We additionally find that accurate rates do not require globally accurate committor models, allowing for efficient kinetics estimation even in a low-data training regime. Requiring little system-specific setup, our approach offers an efficient and broadly generalizable route to binding kinetics, and its shared committor architecture lays crucial groundwork for probing structure--kinetics relationships across ligand series in drug discovery.}


\maketitle

\section{Introduction}\label{introduction}

Molecular simulation play an integral role in drug discovery campaigns. Binding free energy methods, in particular, enable computational chemists to screen large series of related compounds \textit{in-silico} for binding affinity to a target~\cite{choderaAlchemicalFreeEnergy2011,Cournia2017,mueggeRecentAdvancesAlchemical2023,courniaApplicationsFreeEnergyCalculations2024}. However, affinity alone does not fully characterize a protein--ligand complex for drug design. Over the last two decades, recognition has grown that binding kinetics play an equally important part in making good drugs~\cite{tumminoResidenceTimeReceptorLigand2008,copelandDrugTargetResidence2016}. \\

The calculation of binding kinetics continues to present a formidable challenge for computational methods. Because free energy is a state function, its estimation does not require sampling the full unbinding path of a ligand, and can instead rely on thermodynamic tricks such as alchemical cycles~\cite{jorgensenEfficientComputationAbsolute1988}. Kinetic properties like ligand residence time, however, do require full sampling of the slow physical unbinding process. Broadly speaking, there are two families of simulation methods for computing ligand residence times. Biasing methods such as infrequent metadynamics~\cite{tiwaryMetadynamicsDynamics2013,tiwaryKineticsProteinLigand2015} and on-the-fly probability enhanced sampling (OPES)~\cite{invernizziOPESOntheflyProbability2021,rayRareEventKinetics2022} drive the slow unbinding transition by biasing the system's dynamics along a collective variable (CV). If the transition state is kept clear of bias potential, the first passage times from biased dynamics can be rescaled to sample the unbiased residence time with high efficiency. In practice, however, good biasing CVs need to capture all the relevant slow degrees of freedom of the protein--ligand system and unambiguously separate the transition state region. Thus, their design requires extensive physical intuition or iterative data-driven workflows~\cite{rizziRoleWaterHostguest2021,leeCalculatingProteinLigand2024}. Path sampling methods, on the other hand, keep the system's dynamics unbiased, while increasing the sampling of reactive rare-event trajectories by seeding unbiased trajectories from frames around the transition state~\cite{dellagoEfficientTransitionPath1998,Dellago2006,bolhuisTransitionPathSampling2002,bolhuisPracticalConceptualPath2015,Zuckerman2017}. While these methods allow for dynamically unbiased sampling of the transition mechanism and its kinetics, they
can become computationally inefficient if starting configurations are not selected around the transition state. Furthermore, efficiently reweighting the path ensemble to recover thermodynamics and kinetics requires a suitable CV. \\

AI for Molecular Mechanism Discovery (AIMMD) is a recently developed path sampling approach that learns the committor on-the-fly by means of an adaptive learning cycle~\cite{jungMachineguidedPathSampling2023,lazzeriMolecularFreeEnergies2023,lazzeriOptimalRejectionFreePath2025,breebaartUnderstandingMechanismsMolecular2026}.  The committor $p$ is the probability that a given configuration reaches one stable state before another, and is considered the ideal reaction coordinate for thermally activated transitions between two states~\cite{bestReactionCoordinatesRates2005,eTransitionPathTheoryPathFinding2010,berezhkovskiiDiffusionSplittingCommitment2013}. In AIMMD, a neural network model of the committor is trained iteratively on short trajectories. This model then informs the selection of starting configurations in the reactive region for new forward and backward simulations (shooting), which are continued until they commit to a state basin. The learned committor also drives reweighting of the path ensemble, which comprises both shooting chains and equilibrium simulations started in the end states. Although AIMMD path sampling can thus reconstruct, from a single run, unbiased reaction mechanisms as well as thermodynamic and kinetic properties, some key challenges --- shared across path sampling methods --- have so far hindered their application to protein--ligand systems. \\

\begin{figure}[H]
\centering
\includegraphics[width=\textwidth]{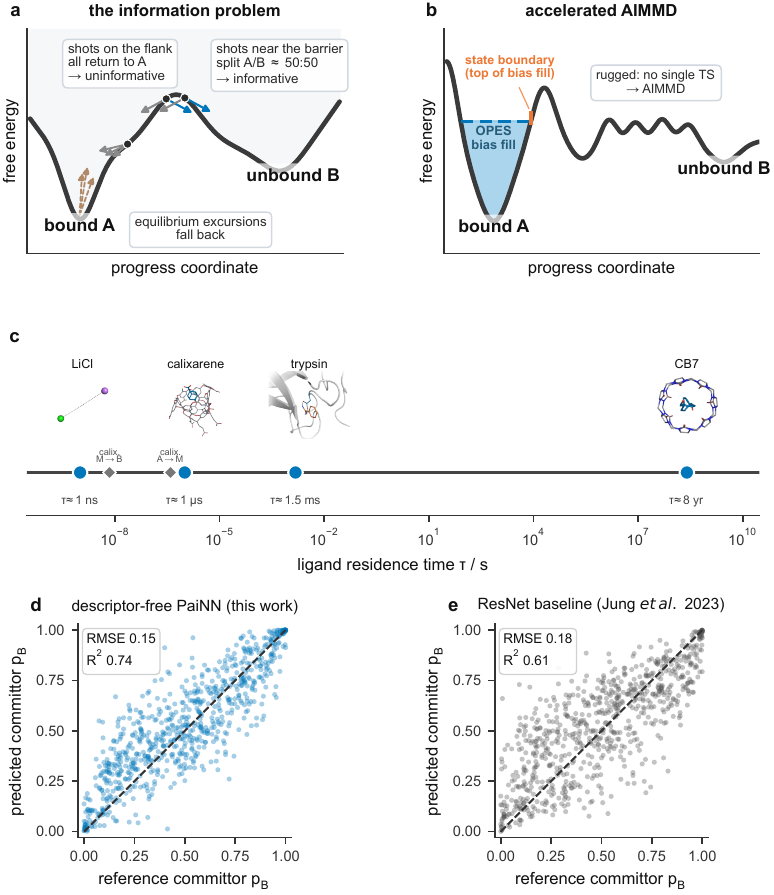}
\caption{\textbf{Accelerated path sampling and descriptor-free committors from graph neural networks} \textbf{a}, Learning the committor from shooting outcomes near the transition region is computationally efficient. On the steep flank of a deep well, shooting outcomes are uninformative, and learning the committor accurately requires much more data, while equilibrium excursions never overlap with informative shooting simulations, 
leading to poor kinetics reweighting convergence. \textbf{b}, A basin-restricted OPES bias partly fills the initial well, allowing for a lifted state boundary and better-converged reweighting;
 the rugged metastable$\rightarrow$unbound region is left to dynamically unbiased AIMMD path sampling. \textbf{c}, One descriptor-free method is
  applied across all systems: LiCl serves as a committor-learning benchmark, while rates are recovered for calixarene (grey diamonds mark the A$\to$M and M$\to$B partial transitions), trypsin and CB7, whose residence times span $\sim$17 orders of magnitude --- from the nanosecond calixarene M$\to$B step to the years-long CB7 release. \textbf{d,e}, Predicted versus brute-force reference committor on the LiCl benchmark from \citet{jungMachineguidedPathSampling2023}, shown as a scatter of individual configurations:
     the descriptor-free PaiNN committor (\textbf{d}: $R^2=0.74$, RMSE $=0.15$) clusters more tightly to the diagonal
      than the hand-featurized ResNet baseline (\textbf{e}: $R^2=0.61$, RMSE $=0.18$), indicating a more accurate committor model.
      }
\label{fig:committor_method}
\end{figure}

Firstly, the neural network committor models commonly used with machine learning-based path sampling methods such as AIMMD were feed-forward networks processing system-specific features. While the inductive bias contained in these feature spaces helps with training well-behaved committor models, the lack of a system-agnostic embedding limits generalization across ligand series, which would be highly desirable in downstream drug discovery applications. Secondly, a more general information problem appears for any path sampling method training committor models on forward/backward shooting outcomes on the slopes of deep free energy wells (figure \ref{fig:committor_method}a).  Shooting outcomes are informative when $p \gtrsim 1/n_\mathrm{sim}$, where $n_\mathrm{sim}$ is the number of shooting outcomes available from configurations close in phase space. Supervised training thus works up to a well depth determined roughly by that number (assuming no overfitting). Deep in the well, however, all shots (of a limited feasible sample) fall into the well, depriving the network of training signal. For obtaining correct kinetics, the committor model must, at least, rank excursions down to the point where equilibrium excursions from in-state simulations can reach and `take over'~\cite{lazzeriOptimalRejectionFreePath2025}. For deep wells, commonly encountered as the first step of ligand dissociation pathways, a gap may remain, leading --- as we show here --- to an overestimation of the rate. Moreover, these two challenges are linked. In certain cases, as shown recently on the CB7-B2 host--guest system, a combination of data augmentation and network regularization along a low-dimensional space of physical features can rescue rate estimation in deep wells~\cite{breebaartUnderstandingMechanismsMolecular2026}. This feature-space inductive bias, however, comes at the cost of a generalizable embedding. \\

Here, we propose a path sampling-based method that can compute high-quality rate estimates at modest computational cost, with minimal system-specific setup and no need for extensive feature engineering. Our strategy follows a `best of both worlds' approach, bringing together recent innovations in the biased sampling and path sampling communities. To this end, we build on the AIMMD adaptive sampling formalism, and incorporate system-agnostic embeddings and static bias potentials into an easily generalizable workflow. Our design choices, as detailed in the following paragraphs, were guided by two principles. We opt for the highest-possible generalizability, and thus did not perform extensive network architecture and hyperparameter optimization, instead using very similar architectures for all systems. We also prioritize sampling parsimony for kinetics convergence, meaning that we intentionally operate in a low-data committor training regime of only a few thousand shooting paths per AIMMD campaign. \\

Inspired by the growing use of graph neural networks (GNNs) for modelling committors, including in iterative workflows for bias-depositing methods~\cite{kangComputingCommittorCommittor2024,kangCommittorsDescriptors2025,arredondoAtomsDynamicsLearning2025,zhangDescriptorFreeCollectiveVariables2024}, we adopt equivariant PaiNNs~\cite{schuttEquivariantMessagePassing2021} as a shared architecture for committor models across all binding systems presented here. These can process Cartesian coordinates directly, thus abolishing the need for feature engineering. At the same time, we draw on an intuition from \citet{rayIntegratingPathSampling2024}, who combined OPES with weighted ensemble (WE) simulations on a host--guest system. Ligand unbinding pathways usually include multi-stage processes. Deep initial wells, for instance from buried hydrophobic contacts or salt bridges, are followed by one or more intermediate states embedded in a more rugged free energy landscape, including protein and ligand rearrangements as well as hydration changes. Biasing methods tend to do well with the first part, which can often be described using simple CVs such as centre-of-mass distances. They may struggle, however, with the second part, where --- in order to keep the transition state bias-free --- complex CVs need to be derived. Path sampling avoids this problem through dynamically unbiased simulations, while struggling, as discussed above, with learning well-behaved committors on the slopes of deep wells. \\

We build on this complementarity of approaches by including in our method, here termed accelerated AIMMD, the option to (partially) flatten deep state wells with OPES biases (figure \ref{fig:committor_method}b). First, we run short OPES simulations in the deep bound state to derive bias potentials, which we then freeze through subsequent AIMMD runs. This allows us to push the AIMMD state boundary further up the deep-well slope, beyond the information-starved region, while keeping a sufficient flux of from-state excursions. Since the bias is designed to be non-zero strictly only within the state boundaries, the transition state ensemble sampled by AIMMD remains dynamically unbiased, and kinetics can be straightforwardly reweighted by applying the formula of \citet{tiwaryMetadynamicsDynamics2013} to the state dwell times (see Methods for details). Taking together the generalizable descriptor-free committors from GNNs and the improved kinetics convergence through flattening the deep state wells with static OPES biases, we show that accelerated AIMMD can robustly recover rates and transition mechanisms across ligand binding systems whose residence times span nearly 17 orders of magnitude, from the nanosecond-scale calixarene--G4 M$\to$B step to the years-long release of CB7-B2. Furthermore, on the real protein--ligand system trypsin--benzamidine, it gives kinetics closely matching experimental data, at moderate computational cost, without the need for system-specific feature engineering or extensive hyperparameter searches.

\section{Results}\label{results}

First, we validated the GNN-based committor models on LiCl dissociation and the within-state OPES biases on the CB7-B2 host--guest system. We then moved to calixarene--G4 host--guest, as a case study for capturing complex, multi-stage and multi-channel binding mechanisms. Lastly, we present converged kinetics and mechanistic insight for the trypsin--benzamidine protein--ligand system.
Kinetics results for all systems are also summarized in table \ref{tab:kinetics}.

\begin{table}[t]
\centering
\caption{\textbf{Cross-system binding kinetics.} Off-rates ($k_\mathrm{off}$, s$^{-1}$) across the host--guest and protein--ligand systems. Standard and accelerated AIMMD values are converged estimates read from the training logs; for the host--guest systems (CB7--B2 and calixarene) they are the mean over three independent production replicates, with the replicate standard deviation in parentheses (in units of the last digit; e.g.\ $3.5(9)$ denotes $3.5\pm0.9$), whereas the trypsin values are single runs. The replicate spread is small throughout ($\le0.3$ dex, i.e.\ within a factor of $\sim$1.8). Unbiased-MD calixarene values are the processed exhaustive-sampling kinetics; the OPES entries are OPES-flooding estimates using 50 transitions. Accelerated AIMMD recovers the reference regime across the systems; for trypsin it lands within a small factor of experiment while standard AIMMD overshoots by $\sim$2 orders of magnitude. ``infeas.'' marks a computationally infeasible brute-force MD reference (CB7--B2, trypsin); ``--'' marks a non-applicable entry (no literature/experimental constant, or acceleration not possible due to lack of suitable, simple CVs). The accelerated-AIMMD entry for the calixarene M$\to$B step is absent because that shallow, diffuse step has no deep initial well to flatten and is handled by standard AIMMD; simple OPES flooding also fails for it. Reference off-rates are cited: CB7-B2 \cite{breebaartUnderstandingMechanismsMolecular2026}, calixarene \cite{debnathComputingRatesUnderstanding2022}; the trypsin literature value is the Ansari 2022 enhanced-sampling estimate \cite{ansariWaterRegulatesResidence2022}, consistent with the experimental $k_\mathrm{off}\approx6\times10^{2}\,\mathrm{s}^{-1}$.}
\label{tab:kinetics}
\fontsize{6.5pt}{7.8pt}\selectfont
\setlength{\tabcolsep}{3pt}
\begin{tabular}{@{}lllllll@{}}
\toprule
System & Standard & OPES& \textbf{Accel.} & Unbiased & Lit. & Expt. \\
 & AIMMD & & \textbf{AIMMD} & MD & & \\
\midrule
CB7-B2 & $3.5(9)\times10^{6}$ & $1.9\times10^{-11}$ & $3.4(11)\times10^{-11}$ & infeas. & $4.0\times10^{-9}$~\cite{breebaartUnderstandingMechanismsMolecular2026} & -- \\
Calix.\ (A$\to$B) & $3.0(8)\times10^{6}$ & $1.5\times10^{6}$ & $4.8(1)\times10^{5}$ & $6.7\times10^{5}$ & $2.4\times10^{6}$~\cite{debnathComputingRatesUnderstanding2022} & -- \\
\quad AM & $6.4(15)\times10^{6}$ & $1.2\times10^{6}$ & $1.8(3)\times10^{6}$ & $2.6\times10^{6}$ & -- & -- \\
\quad MB & $1.2(2)\times10^{8}$ & $5.3\times10^{6}$$^{*}$ & -- & $1.4\times10^{8}$ & -- & -- \\
Trypsin--bzn & $1.2\times10^{5}$ & -- & $1.0\times10^{3}$ & infeas. & $7.0\times10^{2}$~\cite{ansariWaterRegulatesResidence2022} & $6.0\times10^{2}$ \\
\bottomrule
\end{tabular}
\end{table}

\subsection{LiCl --- validating GNN committor models}

We first sought to verify that descriptor-free GNN-based committor models can behave as well as the feed-forward networks on custom features used previously with AIMMD. To this end, we trained out-of-the-box equivariant GVP~\cite{jingEquivariantGraphNeural2021} and PaiNN~\cite{schuttEquivariantMessagePassing2021} models on a LiCl dissociation dataset from AIMMD runs by \citet{jungMachineguidedPathSampling2023}, and compared our results to the ResNet architecture of that publication, which used a 221-dimensional feature space of various order parameters. As shown in figure \ref{fig:ed_licl_detail}, GNN training converges to slightly lower validation-set losses and is notably more data efficient (with PaiNN outperforming GVP on the committor learning task, at equivalent network size). Moreover, the PaiNN committor model is better calibrated than ResNet as measured by comparison to brute-force committor sampling via shooting simulations, using the same training and validation sets (figure \ref{fig:committor_method}e). We thus conclude GNN committor models to be more robust while also outperforming previous feed-forward networks, and use them for all subsequent systems.

\subsection{CB7--B2 --- converging AIMMD kinetics in a deep well}

Next, we applied the GNN-enabled AIMMD workflow to the CB7--B2 host--guest system, which exhibits an exceptionally deep free-energy well of $\approx45~k_BT$ and a guest residence time of several years~\cite{breebaartUnderstandingMechanismsMolecular2026}. Although no ground truth using unbiased MD exists for this system, we picked it as a first test case, since --- with its dominant deep well --- it constitutes a notable challenge for path sampling kinetics methodology, while still avoiding some of the mechanistic complexities of the other covered systems (such as the multiple reaction channels of calixarene). In previous AIMMD work on this system by \citet{breebaartUnderstandingMechanismsMolecular2026}, an off-rate estimate of \qty{4e-9}{\per\second} was recovered, using a 14-dimensional hand-crafted feature space, extensive data augmentation, and regularization with a variational-style loss applied on the low-dimensional feature space. Since a ground-truth reference rate from unbiased MD is not feasible, we first sought to verify the published rate with a standard OPES flooding run and obtained \qty{1.9e-11}{\per\second} (using the host--guest centre-of-mass distance as a one-dimensional CV). These rate estimates from literature AIMMD and our simple OPES flooding setup differ by a factor of $\approx 200$. We have no clear basis here on which to adjudicate between these candidate ground-truth numbers: our simple OPES CV may not have separated the transition state suitably well, there may have been a residual committor inaccuracy deep in the well in the literature AIMMD, or there may have remained --- despite our best efforts --- a difference between the OpenMM-based simulations of \citet{breebaartUnderstandingMechanismsMolecular2026} and our GROMACS setup (or any combination thereof). Therefore, we consider any rate estimate falling in this range comparable to state-of-the-art methodology (and note that the systematic error of standard path sampling on CB7-B2, as shown below, far exceed this residual). \\

\begin{figure}[t]
\centering
\includegraphics[width=\textwidth]{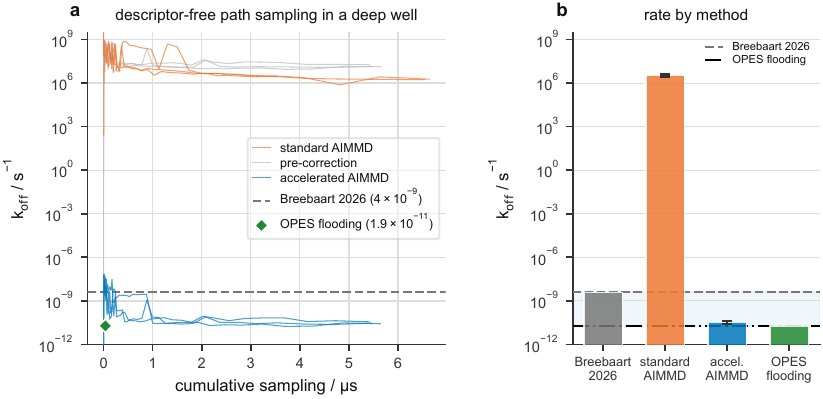}
\caption{\textbf{CB7-B2: descriptor-free path sampling in a deep well.} \textbf{a}, Off-rate estimate versus cumulative 
aggregate sampling (as reported by run-concurrent AIMMD trainer). Standard 
descriptor-free AIMMD (orange) exhibits a pathological regime (triplicate mean $(3.5\pm0.9)\times10^{6}\ \mathrm{s}^{-1}$, $\sim$15 orders above reference),
 whereas accelerated AIMMD (blue; grey shows the pre-bias-correction estimate) quickly drops into and holds the correct regime
  ($3.4\pm1.1\times10^{-11}\ \mathrm{s}^{-1}$). Dashed line, Breebaart 2026
   ($4\times10^{-9}\,\mathrm{s}^{-1}$); dash-dot, our OPES-flooding estimate ($1.9\times10^{-11}\,\mathrm{s}^{-1}$). \textbf{b}, Off-rate
    by method (log scale); accelerated AIMMD lands in the reference band (shaded) while standard AIMMD overshoots by many orders.}
\label{fig:cb7}
\end{figure}

When we ran triplicate  standard AIMMD campaigns of CB7--B2 guest (un)binding with the shared GNN committor models, we consistently found the unbinding rate constant to be overestimated by more than 15 orders of magnitude (figure \ref{fig:cb7}). Despite cumulative simulation time reaching several microseconds in each replicate, no trend towards the correct rate estimate was visible. On the other hand, the accelerated AIMMD campaigns converged to within the expected error on the timescale of 10--100 nanoseconds of cumulative sampling, comparable to pure OPES flooding.  For comparison, we also attempted a (system-specific) array of alternative featurization approaches, neural network architectures and loss function additions, with improvement observed only in rare configurations, and with substantial training instabilities (see figures \ref{fig:ed_negative_results} and \ref{fig:ed_cb7_committor}). While our method can converge escape rates from deep wells with little to no system-specific adaptation, ad-hoc fixes without static bias potentials --- at least in the low-data regime --- appear hard to find and less robust.

\subsection{Calixarene--G4 --- multi-stage ligand unbinding paths}

\begin{figure}[H]
\centering
\includegraphics[width=\textwidth]{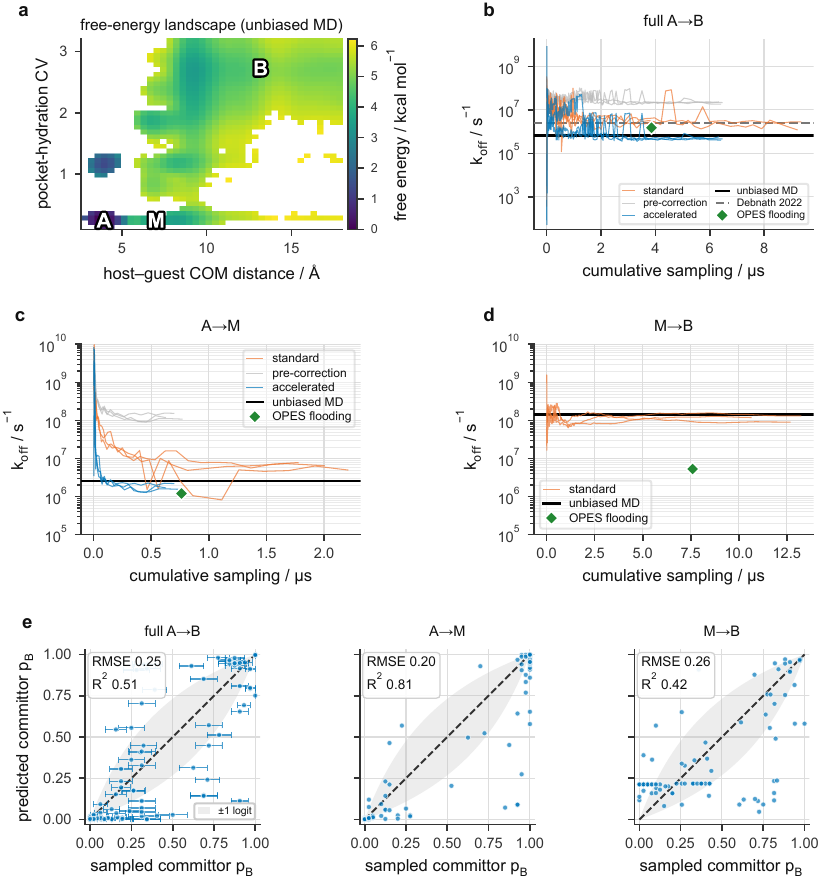}
\caption{\textbf{Calixarene-G4: accelerated AIMMD vs unbiased MD ground truth.} \textbf{a}, Unbiased-MD free-energy
 surface in host–guest COM distance and the pocket-hydration (water-density) CV, with bound (A), metastable (M) and unbound (B) basins highlighted; 
 A and M both sit at low pocket hydration (dry cavity), while B is largely water-filled. \textbf{b}, Full A→B off-rate versus cumulative sampling. 
 Biasing \emph{only} the initial A well with OPES (blue; grey = pre-correction) and letting path sampling handle the M→B escape recovers the 
 unbiased-MD rate (solid black, $6.66\times10^{5}\,\mathrm{s}^{-1}$; accelerated AIMMD $(4.8\pm0.1)\times10^{5}\,\mathrm{s}^{-1}$, triplicate mean $\pm$ s.d.), while standard
 AIMMD converges slower. Dashed grey, Debnath \& Parrinello 2022 (full transition); green diamond, OPES flooding on the full transition with bias constrained to the initial well. \textbf{c,d}, 
 The two sub-steps disentangled: for the deep-well A→M step (\textbf{c}) OPES flooding based on a distance CV is within a small factor of the unbiased $2.57\times10^{6}\,\mathrm{s}^{-1}$,
  whereas for the diffuse M→B step (\textbf{d}; no single transition state) naive distance-CV-based OPES flooding fails ($\sim$30$\times$ low vs $1.43\times10^{8}\,\mathrm{s}^{-1}$);
   AIMMD (accelerated for the A→M step) handles both. \textbf{e}, Predicted versus brute-force sampled committor for the full A→B ($R^2=0.51$), A→M ($R^2=0.81$) and M→B ($R^2=0.42$)
    production committors. Accelerated AIMMD committors where bias could be used, standard
     path sampling where not. All committor calibration panels are the best-converged replicate of the production runs above; shaded band, $\pm1$ logit.}
\label{fig:calixarene}
\end{figure}

To investigate in detail how our method handles more complex, multi-stage ligand unbinding paths typical of real protein--ligand systems, we turned to calixarene--G4 host--guest as a suitable case study. It has previously served as a benchmark for biased enhanced sampling methods and is known as a challenging test case for CV-based methods, owing to the role of hydration in ligand unbinding following exit from the initial deep well, and the presence of two unbinding channels (dry and wet)~\cite{rizziRoleWaterHostguest2021,debnathComputingRatesUnderstanding2022,rayIntegratingPathSampling2024}. First, we established ground-truth reference kinetics and free-energy landscapes by running a total of \qty{160}{\micro\second} of unbiased MD (figure \ref{fig:calixarene}a, and values in table \ref{tab:kinetics}). Our simulations confirm that calixarene host--guest literature kinetics are within a factor of $\approx 3$ of unbiased kinetics and recover the dry and wet pathways previously characterized. For the purposes of our analysis, we distinguish between the bound state A, a (dry) metastable state M, and the unbound state B (see also figure \ref{fig:ed_state_sensitivity}).\\

We then ran triplicate AIMMD campaigns, both on the full transition and the partial A→M and M→B transitions (figures \ref{fig:calixarene}b--d, see also figure \ref{fig:ed_association} for association kinetics). All runs eventually converged to the ground-truth estimate, though in the case of the full and A→M transitions --- that is, those involving the deep initial well --- correct results were only obtained with 1--10 microseconds of cumulative sampling, on the order of the unbiased residence time of this system. By contrast, accelerated AIMMD converged $\approx100$ times faster than the standard variant for A→M, within 100 nanoseconds. We also attempted to estimate kinetics with OPES flooding, using a centre-of-mass distance CV. This worked well for the A→M partial transition, and also gave correct rates for the full transition (although $\approx 10$ times more sampling was required, given we could only bias up to the A→M transition state). We found it impossible, however, to use naive OPES flooding with a simple distance-based CV to accelerate the M→B transition. This partial transition has a complex water-entry mechanism that does not resolve to a single transition state in a simple distance CV (see also the Discussion), highlighting the difficulty of the system (literature applications of biased sampling to these transitions have used elaborate data-driven schemes to identify a suitable biasing CV~\cite{rizziRoleWaterHostguest2021,debnathComputingRatesUnderstanding2022}). \\

Interestingly, we found that correct rates could be obtained in the low-data regime (involving only hundreds to a few thousand shooting paths) by reweighting the path ensemble using committor models that are relatively noisy in the transition region, as seen in committor calibration plots (see figure \ref{fig:calixarene}e, and also the Discussion). For example, on the full transition, our final committor model had an RMSE to brute force sampling of 0.25 and an R$^2$ of 0.51, compared to the RMSE 0.15 and R$^2=$0.74 in the LiCl benchmark of figure \ref{fig:committor_method}d. This prompted us look more closely at how exactly accelerated AIMMD improves kinetics convergence compared to standard AIMMD. Firstly, we found that OPES-enabled AIMMD did not significantly outperform standard AIMMD in terms of committor calibration to brute force sampling (figure \ref{fig:ed_opes_control}a). Close to the transition region (where brute force shooting is informative), committor learning is evidently not significantly improved. Moreover, cross-applying the OPES-sampling-trained committor model to the standard AIMMD path ensemble led to pathological rates comparable to the standard sampling-native committor. Similarly, cross-applying the standard sampling-trained committor on the OPES-sampled path ensemble gave correct rates (figure \ref{fig:ed_opes_control}b). Therefore, committor models trained by OPES-accelerated AIMMD are also not better close to the state boundary. Instead, the information problem of figure \ref{fig:committor_method}a is bypassed. Applying in-state biases allows us to shift the state boundary upwards, out of the regime where information-starved training leads to very inaccurate committor models and overestimated rates. The committor models remain inaccurate in the information-starved region, but are no longer required to be accurate as far down the well as before to obtain correct kinetics. \\

Lastly, we also noticed in both our exhaustive unbiased MD and AIMMD runs that the cavity can accommodate both the G4 ligand and an extra water (as visible in figure \ref{fig:calixarene}a, details and representative snapshot in \ref{fig:ed_qstate}). This is an off-pathway side-state that exchanges with the binding pathway without disturbing the overall kinetics too much. Here, we do not concern ourselves further with it, beyond noting that AIMMD correctly identifies it as part of the unbinding mechanism.

\subsection{Trypsin--benzamidine --- AIMMD kinetics for a protein--ligand system}

\begin{figure}[H]
\centering
\includegraphics[width=\textwidth]{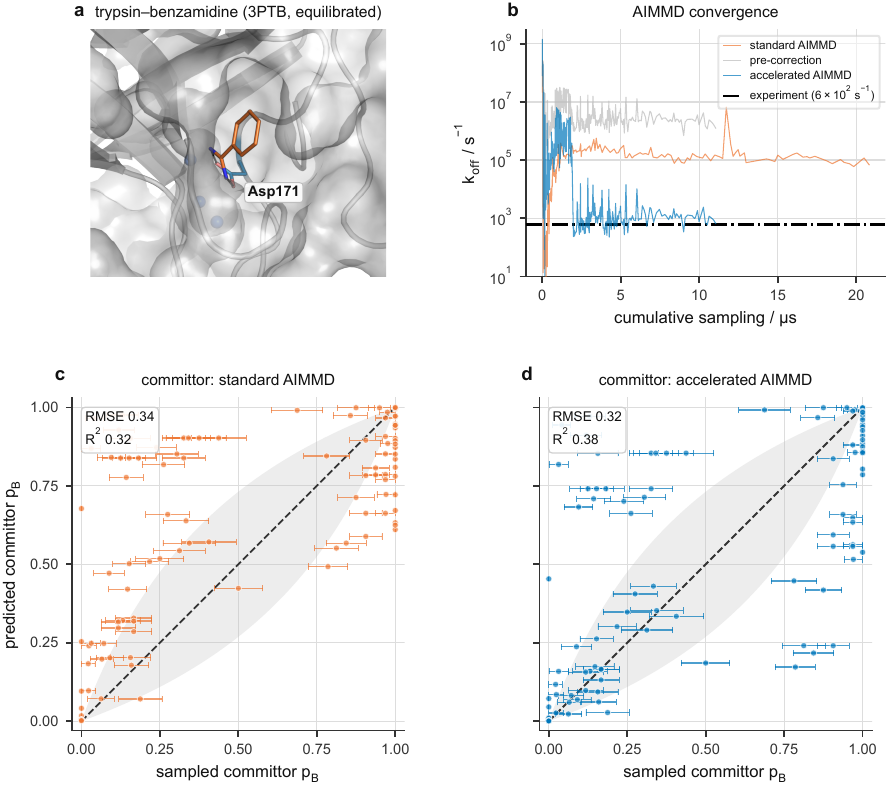}
\caption{\textbf{Trypsin–benzamidine: a realistic drug-like target compared with experimental kinetics.} \textbf{a}, The complex (equilibrated from pdb 3PTB): trypsin (cartoon)
 with benzamidine bound in the S1 pocket and the amidinium–Asp171 salt bridge (black dashes; Asp171 in this box, Asp189 in canonical 3PTB numbering); pale-blue
  spheres are the cavity waters within 5 Å of Asp171. \textbf{b}, Off-rate estimate versus cumulative sampling. A brute-force reference is infeasible (would cost $\sim$3500 GPU-days), so the reference line is the experimental
    $k_\mathrm{off}\approx6\times10^{2}\,\mathrm{s}^{-1}$ ($\tau\approx1.5$ ms), consistent with previous enhanced sampling studies~\cite{tiwaryKineticsProteinLigand2015,ansariWaterRegulatesResidence2022,maityQuantitativePathwayResolvedKinetics2026}. Descriptor-free accelerated AIMMD reaches
     $1.0\times10^{3}\,\mathrm{s}^{-1}$ (a factor of $\sim$1.7 above the experimental $6\times10^{2}\,\mathrm{s}^{-1}$); standard AIMMD overshoots ($1.2\times10^{5}\,\mathrm{s}^{-1}$, $\sim$200$\times$ experiment) even
      with more sampling (both trypsin values are from single runs; see Methods). \textbf{c,d}, Predicted versus brute-force sampled committor on the held-out validation frames for the committor trained
       in the standard run (\textbf{c}, $R^2=0.32$) and in the accelerated/OPES run (\textbf{d}, $R^2=0.38$); both committor fits are mediocre, yet only the accelerated run's recovered
        rate lands near experiment. Shaded band, $\pm1$ logit.}
\label{fig:trypsin}
\end{figure}

Trypsin-benzamidine is a paradigmatic protein--ligand system with millisecond-timescale unbinding kinetics. Benzamidine is bound in a relatively shallow pocket, in a deep initial free energy well associated with a salt bridge to Asp171, and a subsequent complex mechanism involving ligand reorientation and water (figure \ref{fig:trypsin}a). Previous enhanced sampling studies have confirmed that --- with suitable, usually data-driven CV design --- kinetics in line with experimental measurements can be obtained from biased~\cite{tiwaryKineticsProteinLigand2015,ansariWaterRegulatesResidence2022} or weighted ensemble simulations~\cite{maityQuantitativePathwayResolvedKinetics2026}. Without involved CV design, however, simple methods like OPES flooding using distance CVs are no longer feasible on this system, highlighting the continuing scarcity of kinetics methods that do not require extensive system-specific setup. \\

Here, we apply our GNN-enabled AIMMD workflow, with and without OPES acceleration of the initial salt bridge breaking, to trypsin--benzamidine (figure \ref{fig:trypsin}b). Accelerated AIMMD converges to correct kinetics at around 2 microseconds of cumulative sampling ($k_\mathrm{off}\approx1.0\times10^{3}\,\mathrm{s}^{-1}$, a factor of $\sim$1.7 above the experimental off-rate), whereas standard AIMMD overshoots by $\approx 2$ orders of magnitude ($\approx1.2\times10^{5}\,\mathrm{s}^{-1}$) even with substantially more sampling, mirroring the deep-well behaviour seen for CB7-B2. We would like to stress that this improved convergence is not due to the in-state OPES bias driving spontaneous transitions in the free A-state simulations that form part of the AIMMD campaign. No such transitions are observed, and accelerated AIMMD does not collapse to OPES flooding, which remains infeasible on this system with simple CVs. We note that committor calibrations of the AIMMD models against brute force sampling do contain information, but are again noisy (see figure \ref{fig:trypsin}c--d, and the Discussion). Nonetheless, the method delivers on-point kinetics in the low-data committor training regime, with reasonable sampling requirements and little system-specific setup.

\section{Discussion}\label{sec3}

In this paper, we showcased a method for computing protein--ligand binding kinetics that combines the `best of both worlds' from biased sampling and path sampling. It delivers robust kinetics on systems spanning 17 orders of magnitude in residence time, with high sampling efficiency, little system-specific setup and no extensive feature engineering or data-driven CV design. This is made possible by two key additions to the AIMMD path sampling framework --- GNN committor models and the application of bias potential within deep-well states. The result is an easily generalizable protein--ligand kinetics method with fast convergence. \\

\begin{figure}[p]
\centering
\includegraphics[width=0.82\textwidth]{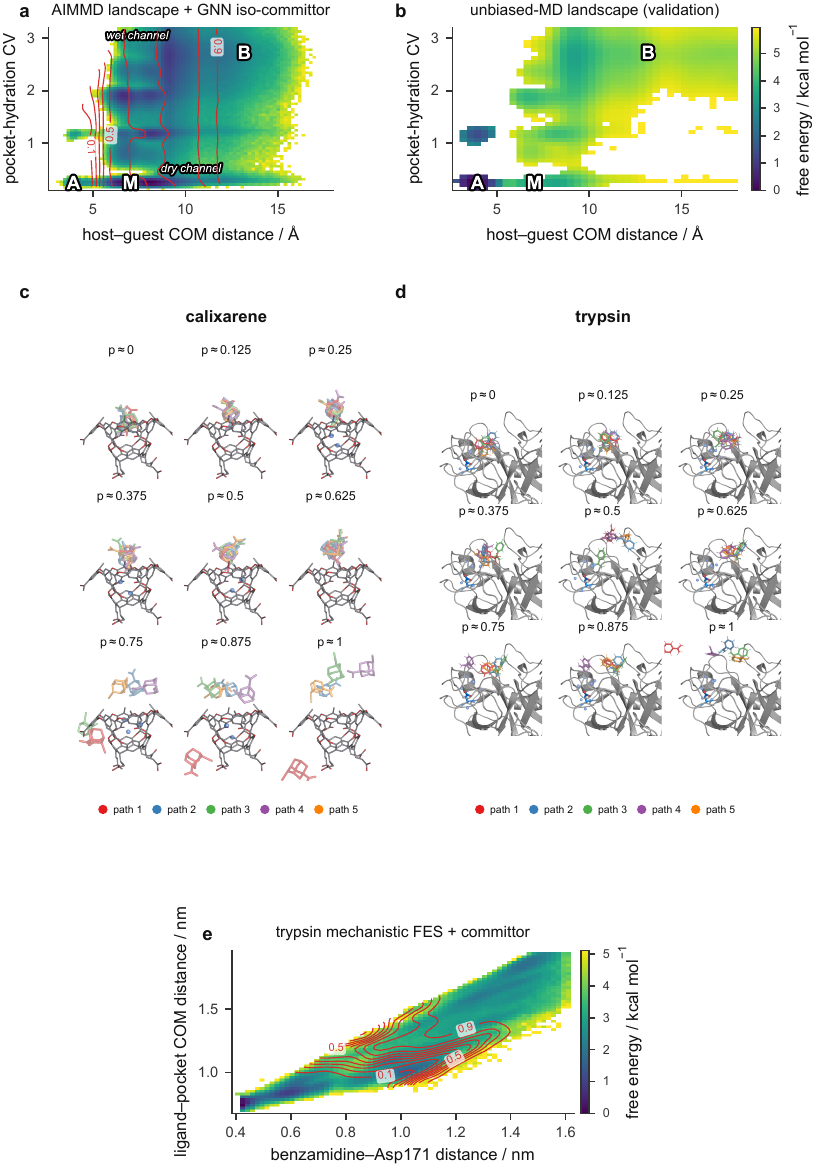}
\caption{\textbf{Molecular mechanism of (un)binding.} \textbf{a}, Calixarene-G4 landscape from the AIMMD full-transition ensemble (sampling free energy,
 $-RT\ln$ density) in host–guest COM distance and the pocket-hydration (water-density) CV, overlaid with red iso-committor contours ($p=0.1$–$0.9$,
  0.1 spacing) from the descriptor-free production committor; A/M/B basins marked and the wet/dry hydration channels labelled. \textbf{b}, The unbiased-MD
   free-energy surface on the same axes and colour scale, validating the basins. \textbf{c}, Calixarene unbinding resolved by committor:
    each of the nine panels fixes a committor value ($p=0$–$1$ in steps of $0.125$) and overlays the guest from five randomly-selected transition paths
     (distinct colours), each the frame whose predicted committor is closest to that value; the host is aligned and the camera fixed across panels,
      revealing both the guest's migration as $p$ increases and the path-to-path spread at fixed commitment.
       Binding-site waters from one representative path are drawn as pale-blue spheres, showing how the cavity wets as the guest departs.
       \textbf{d}, The same committor-resolved five-path overlay for benzamidine leaving the trypsin S1 pocket (protein-aligned), with S1-pocket waters likewise shown as pale-blue spheres. \textbf{e}, Accelerated-AIMMD
        reweighted trypsin free-energy surface in the benzamidine–Asp171 salt-bridge distance and the ligand–pocket COM distance, with red iso-committor contours
         from the production committor — the salt-bridge rupture gates the exit.}
\label{fig:mechanism}
\end{figure}

Moreover, since AIMMD keeps the transition path ensemble dynamically unbiased (and reweighting recovers the relative probabilities of paths), it delivers mechanistic information at the same time, without no extra sampling required. As shown in figure \ref{fig:mechanism}a--b, AIMMD produces a free energy landscape of the calixarene--G4 unbinding process that closely matches the shape of the unbiased MD reference, including capturing the dry and wet channels (although the scale of the free energy landscape is off by $\approx 1 \text{kcal}~\text{mol}^{-1}$, owing to a slight overstabilization of the reactive region). Interestingly, overlaying this landscape with the GNN committor predictions reveals that the host--guest centre-of-mass distance captures most of the variation in committor values, so that although the dry and wet channels are clearly distinct, the unbinding progress coordinate is approximately shared (yet orthogonal to their interconversion). As figure \ref{fig:mechanism}c further shows, at late stages of the unbinding process (from  $p \approx 0.75$), the G4 ligand can associate freely and without directional constraint with the calixarene host --- a finding not contained in previous enhanced sampling studies (that have used funnel restraints). \\

Similarly, in figure \ref{fig:mechanism}d--e, we show a mechanistic interpretation of the trypsin-benzamidine unbinding path ensemble. As with calixarene, the ligand position and orientation become more structurally diverse at higher committor values, though in this case, loose association with the protein alone is not sufficient to push the committor significantly below 1. As the free energy landscape superimposed with committor isolines reveals, unbinding proceeds in a two-step process: first, the benzamidine--Asp171 salt bridge breaks, but this alone does not commit a configuration hovering at similar centre-of-mass distance in the pocket to full unbinding. Instead, a more complex pathway (with several sub-transition states that are only partially resolved in the CV space used for this projection) is captured in how the committor rises with increasing centre-of-mass distance between 1.0 and \qty{1.3}{\nano\meter}. While the design of a full set of physical order parameters that could resolve the fine detail of this free energy landscape is beyond the scope of this kinetics-focused paper, the non-trivial nature of CV design for protein--ligand systems --- rendered unnecessary by our method --- stands highlighted once again. \\

Despite the well-converged and robust kinetics presented for all ligand binding systems studied here, we must note that our estimated committor models were only modestly accurate. In fact, an important conclusion appears to be that, within the regime of moderate committor calibration that we probe, the correlation of the committor to brute-force shooting does not by itself determine the quality of the kinetics results (see also the overview in figure \ref{fig:ed_committor_vs_rate}). A catastrophically miscalibrated committor would of course spoil the rate; the point is rather that, across the moderate-$R^2$ range spanned by our systems, where the state boundary is placed matters more than the committor's transition-region $R^2$. Rate estimate quality relates to how well a committor model extrapolates along steep free-energy slopes. This is determined in large part by network and training inductive biases. Problems can however be circumvented by the accelerated AIMMD method presented here, which pushes the state boundary to regions in configuration space where the training is smoother. We further note that all AIMMD runs presented here lie, with just a few thousand shooting paths, in a low data regime. Fitting more accurate committor models (and thus, potentially, converging more accurate mechanistic representations) requires substantially more training data, and lies beyond our scope, which focused primarily on rate convergence. An upshot is that converging rates with our method is very sampling efficient, comparable to pure OPES flooding with simple CVs, where it is feasible, while never requiring the design of biasing CVs for the whole transition in cases where it is not. \\

Lastly, we wish to highlight once again the general nature of our method. As shown in figure \ref{fig:ed_architecture}, the neural network architecture and training procedure were deliberately only minimally adapted for each system, primarily for technical reasons to maximize training efficiency given GPU memory constraints. We recommend that users collect a small data set (hundreds of paths) with a default expressive model and small batch size, and then tune some core hyperparameters to make training as efficient as possible, given the constraints of their specific compute architecture. In our trials, we found none of these parameters to substantially affect the scientific quality of the results. Similarly, we adopted very simple end-state definitions based on one-dimensional distance CVs (see Methods) and never needed to revise them, except to separate out the M metastable state for calixarene for investigative purposes. The only genuinely user-defined and (somewhat) system-specific parameters of the accelerated AIMMD method were the OPES bias-flattening CV and OPES hyperparameters. For the host-guest systems, we found naive one-dimensional CVs to work well, and only tuned the OPES maximum bias based on a simple trial series (so that no spontaneous out-of-well transitions were driven on the sub-nanosecond timescale). For trypsin-benzamidine, we found in an initial trial that a one-dimensional distance CV cannot in this case cleanly capture the protein--ligand salt-bridge state, so we added the salt bridge distance as a second CV into the OPES bias (while the state definition remained in the simple centre-of-mass distance CV). While all OPES protocol design choices therefore remain standard  and no elaborate CV design is needed (since only the first well needs to be partially filled), we do recommend that users conduct small preliminary experiments for new systems with pure OPES simulations starting from the bound state. \\

In conclusion, we have shown that our path sampling-based method, based on AIMMD and incorporating GNN committor models and within-state static bias potentials, is a robust and easily generalizable method for investigating protein--ligand binding systems. Alongside mechanistic and thermodynamic insight, the method recovers converged unbinding kinetics with high sampling efficiency, traditionally a limitation of path sampling methods. We anticipate this to beneficially complement established CV-based methods for binding kinetics, and hope that its ease of use and general nature can help researchers tackle those realistic binding problems which so far have required extensive system-specific CV design. Moreover, the system-agnostic nature of the embedding used will in the future allow for extensions to describe multiple systems with one trained committor model, thus laying crucial groundwork for efficiently probing structure--kinetics relationships across protein--ligand series in drug discovery applications. \\ 

The code for running accelerated AIMMD is available as a python package at \href{https://github.com/covinolab/AIMMD-Lab}{https://github.com/covinolab/AIMMD-Lab}. Full input files for reproduction of all simulations, and key results are compiled in the supplementary data, \href{https://doi.org/10.5281/zenodo.21396356}{https://doi.org/10.5281/zenodo.21396356}.

\section{Acknowledgements}\label{sec4}

We gratefully acknowledge funding from the Goethe University, FIAS and the Johanna Quandt foundation, as well as compute resources from the CSC Goethe cluster and the JSC JUPITER cluster. We would like to thank Gianmarco Lazzeri for training, helpful discussions on AIMMD kinetics reweighting and his work on the AIMMD codebase. We also thank Aly Valiev for helpful discussions on GNNs for committor learning. Large Language Models (Claude Opus 4.8, Claude Code) were used during this research, to assist compilation of this manuscript, and to refine the text for ease of reading.

\section{Methods}\label{sec5}

\subsection{Descriptor-free committor path sampling with rejection-free reweighting}

We sampled reactive (un)binding transition pathways and from-state excursions with Artificial Intelligence for Molecular Mechanism Discovery (AIMMD) path sampling \cite{jungMachineguidedPathSampling2023}, in the rejection-free formulation of Lazzeri et al.\ \cite{lazzeriOptimalRejectionFreePath2025}. Rather than sampling the transition-path distribution directly, we sample the distribution of \emph{shooting points} with a selection bias equal to the inverse of the free energy along a reaction coordinate; because the target and selection biases coincide, every shooting move is accepted (rejection-free) while detailed balance is preserved, so the reactive region is seeded uniformly. The optimal reaction coordinate is the logit-committor $\lambda(x)=\ln[p(x)/(1-p(x))]$, and the committor $p$ is learned online by maximising the likelihood of the observed shooting outcomes. \\

Unbiased free energies, mechanisms and rate constants are obtained by reweighting the sampled paths to equilibrium following the rejection-free path-sampling scheme \cite{lazzeriOptimalRejectionFreePath2025,lazzeriMolecularFreeEnergies2023}: each path is assigned a continuous shooting-interface coordinate, a frame-density correction removes over-representation at the shooting interface, and excursions from each basin are reweighted by matching the sampled crossing histograms to the equilibrium crossing probabilities, with the two-state flux balance fixing the kinetics through a maximum-likelihood estimator. From the resulting reweighted path ensemble we read the off-rate $k_\mathrm{off}$; rates quoted here are taken directly from the AIMMD training logs and never re-fitted. Log rates are per stored frame and are converted to $\mathrm{s}^{-1}$ using each run's integration time step and output stride ($\Delta t \times$ \texttt{nstxout-compressed}).

\subsection{Equivariant-GNN committor}

The committor is represented by a single descriptor-free, E($n$)-equivariant message-passing neural network of the PaiNN type \cite{schuttEquivariantMessagePassing2021}, acting on the atomic graph of heavy atoms within a fixed radial cutoff of the ligand; no hand-crafted order parameters are used as feature spaces for the committor models. The same architecture is applied to all four systems, with only minor per-system adjustment of the network width/depth and training-set size to efficiently train given GPU memory constraints, see figure \ref{fig:ed_architecture}. The network outputs a single committor logit per configuration and is trained by shooting-outcome cross-entropy within the AIMMD loop. The graph comprises the ligand or guest heavy atoms together with all heavy atoms within 6--8~\AA{} of them, with element-specific embeddings for the species present (H, C, N, O; additionally S and Cl for trypsin, and Li/Cl for the LiCl benchmark). Per-system architecture and training hyperparameters are listed in Table~\ref{tab:committor}.

\subsection{Basin-restricted OPES acceleration and reference rates}

To accelerate the sampling of rare escapes we add an on-the-fly probability-enhanced-sampling (OPES) bias \cite{invernizziRethinkingMetadynamicsBias2020,invernizziOPESOntheflyProbability2021} that is \emph{restricted to the bound basin}, within AIMMD's state cutoff: using PLUMED's \texttt{EXCLUDED\_REGION}, the bias is deposited only inside the initial well (in the host--guest and protein--ligand systems, on the ligand--host/pocket centre-of-mass distance, and for trypsin additionally on the benzamidine--Asp171 salt-bridge distance). The purpose of this restriction is not merely to speed up escapes: flattening the deep bound well lets us move the \emph{state boundary further up the well}, away from the free-energy minimum and towards the barrier region, while leaving the boundary and everything beyond it bias-free. This shifted boundary is what makes the committor learnable and the reweighting stable, as shown in the main text. First, shooting points are then drawn from near the transition region, where the committor actually varies, instead of from the deep-well flank, where every shot commits back to the bound state and carries no committor information; the network therefore learns from informative shots (cf.\ Fig.~\ref{fig:committor_method}a, the committor ``information problem''). Second, the excursions that the path reweighting must account for become shorter and better sampled, so the crossing-probability matching which yields the rate is far less sensitive to rare, poorly-sampled deep-well fluctuations---improving the stability of the reweighted rate estimate across replicates. Path sampling then only has to resolve the remaining barrier(s) beyond the boundary. Per-system bias collective variables, barriers ($10$--$100\,\mathrm{kJ\,mol^{-1}}$), bias factors and deposition windows, together with the parameters of the independent OPES-flooding reference runs, are collected in Table~\ref{tab:enhanced}. \\

The converged bias is frozen to a static function $V(s)$ of the restricted collective variable(s) $s$ and reweighted out following Tiwary and Parrinello \cite{tiwaryMetadynamicsDynamics2013}. Equilibrium (unbiased) expectations of any observable $O$ are recovered from the biased ensemble $\langle\cdot\rangle_V$ by importance reweighting,
\begin{equation}
\langle O\rangle_0 \;=\; \frac{\big\langle O\,e^{\beta V(s)}\big\rangle_V}{\big\langle e^{\beta V(s)}\big\rangle_V},\qquad \beta=\frac{1}{k_\mathrm{B}T},
\label{eq:staticrw}
\end{equation}
so each frame of a biased path carries a Boltzmann weight $w=e^{\beta V(s)}$. These per-frame weights are folded directly into the rejection-free reweighted path ensemble introduced above: the basin crossing-probability histograms $m_A(\lambda),m_B(\lambda)$ and frame-density corrections are accumulated with $w$ rather than unit weight, so the reweighted ensemble --- and hence the free-energy projections, mechanism and off-rate $k_\mathrm{off}$ --- is unbiased even though the sampling was accelerated. Because $V$ is identically zero on the barrier and beyond (\texttt{EXCLUDED\_REGION}), the transition state and all reactive crossings occur at their physical rates; only the residence time in the bound well is compressed. This same requirement---that the bias never perturb the transition state---underlies the independent OPES-flooding / infrequent-metadynamics reference rate \cite{rayRareEventKinetics2022,tiwaryMetadynamicsDynamics2013}: for a bias confined below the transition state, physical time dilates relative to biased time by the acceleration factor
\begin{equation}
t_\mathrm{phys}=\int_0^{t_\mathrm{bias}} e^{\beta V(s(t'))}\,\mathrm{d}t' \equiv \alpha\,t_\mathrm{bias},\qquad \alpha=\big\langle e^{\beta V(s)}\big\rangle_t ,
\label{eq:accel}
\end{equation}
where $\alpha$ is the running time-average of $e^{\beta V}$; the unbiased mean first-passage time $\tau=\langle t_\mathrm{phys}\rangle$ over independent basin-restricted escapes then gives $k_\mathrm{off}=1/\tau$. Equation~\eqref{eq:staticrw} supplies the frozen-bias weights used inside AIMMD, while Eq.~\eqref{eq:accel} yields the independent flooding reference. \\

All molecular dynamics were run with GROMACS \cite{Abrahams2015} and PLUMED \cite{Tribello2014}, with a 2~fs time step, LINCS-constrained hydrogen bonds, particle-mesh Ewald electrostatics and a velocity-rescaling thermostat at 300~K.

\subsection{Systems, force fields and equilibration}

Composition and simulation-protocol details for the four systems are collected in Table~\ref{tab:systems}, state and committor-network settings in Table~\ref{tab:committor}, and initial-path and enhanced-sampling parameters in Table~\ref{tab:enhanced}; system-specific provenance and equilibration are described below.

\paragraph{LiCl dissociation.} A single Li$^+$/Cl$^-$ pair in 370 TIP3P waters (Joung--Cheatham monovalent-ion parameters; 1112 atoms) serves purely as a committor-learning benchmark. The path-sampling shooting data and the held-out committor-validation set were taken and re-processed from the AIMMD study of \citet{jungMachineguidedPathSampling2023}; a matching box was additionally built in OpenMM, converted to GROMACS, and run (steepest-descent minimization, 200~ps NVT, 1~ns NPT, then 500~ns unbiased MD) to confirm the setup. The committor models --- the 221-feature ResNet baseline of \citet{jungMachineguidedPathSampling2023} against descriptor-free GVP~\cite{jingEquivariantGraphNeural2021} and PaiNN~\cite{schuttEquivariantMessagePassing2021} graph networks --- were trained offline on 26{,}001 shooting points and scored against 763 brute-force committor estimates. For path sampling the bound and dissociated states were a Li--Cl distance below 2.3~\AA{} and above 4.8~\AA{}, respectively.

\paragraph{Cucurbit[7]uril--B2 (CB7-B2).} The host--guest complex was parameterized with GAFF and AM1-BCC partial charges (AmberTools), solvated in a 3.58~nm cubic TIP3P box with no added ions (charge-neutral; 4491 atoms), and converted to GROMACS. After steepest-descent minimization, 200~ps NVT and 1~ns NPT (C-rescale barostat) equilibration, path sampling was run in NVT with 1.2~nm interaction cutoffs. Initial unbinding paths were generated by adiabatic-bias MD (ABMD) on the host--guest centre-of-mass (COM) distance (harmonic restraint moving to 1.75~nm, force constant $800\,\mathrm{kJ\,mol^{-1}\,nm^{-2}}$); the bound state was defined below 0.42~\AA{} and the unbound state above 14~\AA{} in that COM distance. The basin-restricted OPES bias (BARRIER $100\,\mathrm{kJ\,mol^{-1}}$, bias factor 40) was accumulated over a short run, frozen, and replayed with deposition confined to the bound well ($d_\mathrm{COM}<0.40$~nm) through \texttt{EXCLUDED\_REGION}. An independent OPES-flooding reference (BARRIER $115\,\mathrm{kJ\,mol^{-1}}$, bias confined below 0.50~nm) gave a mean-first-passage off-rate of $1.9\times10^{-11}\,\mathrm{s^{-1}}$ from 25 independent escapes (37.7~ns aggregate); the literature reference off-rate is the deep-well estimate of Breebaart et al.\ \cite{breebaartUnderstandingMechanismsMolecular2026}.

\paragraph{Calixarene (octa-acid)--G4.} The methyl octa-acid host and G4 guest (GAFF, TIP3P, nine neutralising Na$^+$; 6533 atoms in a 4.03~nm cubic box) derive from Rizzi et al.\ \cite{rizziRoleWaterHostguest2021} and, following that work, were equilibrated and run entirely in NVT (steepest-descent minimization, then production; velocity-rescaling thermostat, 1.1~nm cutoff). Extensive unbiased MD --- 64 trajectories totalling $\approx160\,\mu$s --- provides a brute-force kinetic ground truth for the A$\to$B transition, via state-to-state mean-first-passage event counting. All initial transition paths were spontaneous unbinding events harvested from this unbiased MD; no adiabatic-bias MD was required. States are defined on the guest--host COM distance (bound A below 4.1~\AA{}, unbound B above 15~\AA{}); the dry metastable state M, at the dry-cavity free-energy minimum beyond the bound well, additionally requires an empty cavity (no water oxygen within 4~\AA{} of the host COM), because at a given COM distance the cavity is degenerate between a dry (metastable) and a water-filled (reactive) configuration, so a distance-only boundary cannot separate the two. The bound-well OPES bias (BARRIER $10\,\mathrm{kJ\,mol^{-1}}$, bias factor 4) was derived over a short run and then frozen, with deposition confined to the 0.30--0.44~nm window (\texttt{EXCLUDED\_REGION}); in the OPES-flooding reference runs, harmonic lower-wall restraints on four host torsions and on the COM distance prevented unphysical rotation of the host aromatic rings as the guest was pushed outward. The OPES-flooding references (BARRIER $20\,\mathrm{kJ\,mol^{-1}}$, tuned to give sub-nanosecond to few-nanosecond exits; 50 independent escapes per step) reproduce the A$\to$M and full A$\to$B rates but fail for the diffuse M$\to$B step, whose water-entry mechanism does not resolve to a single transition state in a distance CV.

\paragraph{Trypsin--benzamidine.} We found the bound pose of the Ansari et al.\ \cite{ansariWaterRegulatesResidence2022} box unstable in our hands (it ships without an equilibration protocol, and a cold start under a constant-pressure barostat led to spontaneous ligand ejection on a microsecond timescale), so we rebuilt the complex from PDB 3PTB with AMBER ff14SB (protein), GAFF2 (ligand) and two-stage RESP (HF/6-31G$^{*}$) benzamidinium charges in TIP3P. The benzamidinium was modelled in its $+1$ state (the system neutralized with nine Cl$^-$), the structural Ca$^{2+}$ and all six disulfide bonds were retained, the catalytic histidine was assigned its $\delta$-tautomer, and 62 crystallographic pocket waters were seeded (31{,}781 atoms in a $6.84\times6.36\times7.29\,\mathrm{nm}$ box). Equilibration proceeded through steepest-descent minimization, heavy-atom-restrained NVT ($1000\,\mathrm{kJ\,mol^{-1}\,nm^{-2}}$ restraints; a 1~ns stage then a 5~ns solvent-fill), unrestrained NVT (1~ns), NPT (1~ns, C-rescale barostat) and an unbiased stability screen, with a 1.1~nm cutoff; this ladder yielded a stable bound complex with the amidinium--Asp171 salt bridge intact ($\approx0.39$~nm). The state-defining CV is the Asp171(C$\gamma$)--benzamidine(amidinium C) salt-bridge distance (bound A below 4.1~\AA{} in the lowered-boundary run used here, 4.8~\AA{} by default; unbound B above 15.8~\AA{}). Initial unbinding paths were generated by ABMD on the benzamidine--pocket COM distance (moving to 2.6~nm, force constant $1000\,\mathrm{kJ\,mol^{-1}\,nm^{-2}}$), terminated when the salt bridge exceeded 1.75~nm. Because a one-dimensional distance CV does not disambiguate the salt-bridge state, the acceleration OPES bias acted in two dimensions (COM and salt-bridge distances; BARRIER $25\,\mathrm{kJ\,mol^{-1}}$, bias factor 10) with deposition confined to the bound region ($d_\mathrm{COM}<0.85$~nm and $d_\mathrm{sb}<0.47$~nm), while the state definition itself remained on the single salt-bridge CV. The experimental off-rate, $k_\mathrm{off}\approx6\times10^{2}\,\mathrm{s}^{-1}$, is reproduced by the enhanced-sampling estimate of Ansari et al.\ \cite{ansariWaterRegulatesResidence2022}. \\

All input files, force fields, equilibration protocols, PLUMED scripts, trained committor networks, committor-validation sets and convergence logs needed to reproduce these results are provided as Supplementary Data at \href{https://doi.org/10.5281/zenodo.21396356}{https://doi.org/10.5281/zenodo.21396356}. The AIMMD implementation is available at \href{https://github.com/covinolab/AIMMD-Lab}{https://github.com/covinolab/AIMMD-Lab}.

\begin{table}[t]
\centering
\caption{\textbf{System composition and simulation protocol.} Per-system box, solvent, ions and atom counts, with the common GROMACS settings. Path-sampling production is run in NVT (fixed box) to keep two-way shooting time-reversible; equilibration includes an NPT stage (C-rescale barostat) for every system except calixarene, which follows the NVT protocol of Rizzi et al.~\cite{rizziRoleWaterHostguest2021}. All small molecules use GAFF/GAFF2; ``JC'' denotes the Joung--Cheatham monovalent-ion parameters and ``Rizzi 2021'' the charges shipped with that reference.}
\label{tab:systems}
\footnotesize
\setlength{\tabcolsep}{5pt}
\begin{tabular}{@{}lcccc@{}}
\toprule
 & LiCl & CB7--B2 & Calix.--G4 & Trypsin--bzn \\
\midrule
Force field       & JC ions & GAFF & GAFF & ff14SB/GAFF2 \\
Ligand charges    & ---     & AM1-BCC & Rizzi 2021 & RESP \\
Water model       & \multicolumn{4}{c}{TIP3P} \\
Box (nm)          & $2.25^3$ & $3.58^3$ & $4.03^3$ & $6.84\times6.36\times7.29$ \\
Water molecules   & 370 & 1445 & 2100 & 9511 \\
Counter-ions      & Li$^+$/Cl$^-$ & none & 9 Na$^+$ & 9 Cl$^-$, 1 Ca$^{2+}$ \\
Total atoms       & 1112 & 4491 & 6533 & 31{,}781 \\
Time step         & \multicolumn{4}{c}{2 fs} \\
Thermostat        & \multicolumn{4}{c}{velocity-rescaling, 300 K} \\
Electrostatics    & \multicolumn{4}{c}{particle-mesh Ewald} \\
Constraints       & \multicolumn{4}{c}{LINCS, hydrogen bonds} \\
Nonbonded cutoff  & 1.1 nm & 1.2 nm & 1.1 nm & 1.1 nm \\
\bottomrule
\end{tabular}
\end{table}

\begin{table}[t]
\centering
\caption{\textbf{State definitions and descriptor-free PaiNN committor.} State boundaries are given in the state-defining collective variable (CV) of each system (distances with the minimum-image convention). One PaiNN network is trained per system; the loss is the shooting-outcome (Bayesian binomial) log-likelihood optimized with Adam. LiCl values are the offline architecture benchmark; the calixarene entry is the full-transition committor (the A$\to$M and M$\to$B sub-transition networks used 3 layers, 32 channels, 10 bases and a 3~\AA{} cutoff). CB7's tight bound cutoff reflects the guest sitting near the host centre of mass; the trypsin state CV is the Asp171(C$\gamma$)--amidinium-C salt-bridge distance.}
\label{tab:committor}
\footnotesize
\setlength{\tabcolsep}{5pt}
\begin{tabular}{@{}lcccc@{}}
\toprule
 & LiCl & CB7--B2 & Calix.--G4 & Trypsin--bzn \\
\midrule
State CV        & Li--Cl dist. & COM dist. & COM dist. & salt-bridge dist. \\
Bound A         & $<2.3$~\AA & $<0.42$~\AA & $<4.1$~\AA & $<4.1$~\AA \\
Unbound B       & $>4.8$~\AA & $>14$~\AA & $>15$~\AA & $>15.8$~\AA \\
Metastable M    & --- & --- & 4.1--8.5~\AA, dry & --- \\
\midrule
Message layers  & 3 & 3 & 2 & 3 \\
Hidden channels & 64 & 32 & 16 & 32 \\
Radial bases    & 10 & 10 & 6 & 10 \\
Graph cutoff    & 6~\AA & 3~\AA & 4~\AA & 4~\AA \\
Dropout         & \multicolumn{4}{c}{0.1} \\
Learning rate   & $10^{-3}$ & $5\times10^{-4}$ & $8.6\times10^{-4}$ & $10^{-4}$ \\
Batch size      & 128 & 16 & 32 & 32 \\
Epochs          & 50 & 2000 & 3300 & 3000 \\
\bottomrule
\end{tabular}
\end{table}

\begin{table}[t]
\centering
\caption{\textbf{Initial-path and enhanced-sampling parameters.} Initial reactive paths came either from spontaneous transitions in unbiased MD or from adiabatic-bias MD (ABMD; moving-restraint target and force constant $\kappa$ listed). The basin-restricted OPES acceleration bias and the independent OPES-flooding reference are OPES\_METAD potentials on the listed CV(s), deposited only within the stated bound-basin window (\texttt{EXCLUDED\_REGION}); the acceleration bias is additionally frozen and replayed. BARRIER in kJ\,mol$^{-1}$, windows in nm, $\kappa$ in kJ\,mol$^{-1}$\,nm$^{-2}$; SIGMA was adaptive throughout. For trypsin the two deposition-window values are the COM and salt-bridge bounds of the 2D bias.}
\label{tab:enhanced}
\footnotesize
\setlength{\tabcolsep}{5pt}
\begin{tabular}{@{}lcccc@{}}
\toprule
 & LiCl & CB7--B2 & Calix.--G4 & Trypsin--bzn \\
\midrule
Initial paths          & unbiased & ABMD & unbiased & ABMD \\
ABMD target / $\kappa$  & --- & 1.75 / 800 & --- & 2.6 / 1000 \\
\midrule
OPES bias CV(s)        & --- & COM & COM & COM, salt-bridge \\
BARRIER                & --- & 100 & 10 & 25 \\
Deposition window      & --- & $<0.40$ & 0.30--0.44 & $<0.85$, $<0.47$ \\
Bias factor            & --- & 40 & 4 & 10 \\
\midrule
Flooding BARRIER       & --- & 115 & 20 & --- \\
Flooding window        & --- & $<0.50$ & $<0.43$ & --- \\
Flooding escapes       & --- & 25 & 50/step & --- \\
\bottomrule
\end{tabular}
\end{table}

\backmatter

\clearpage

\bibliography{library}

\clearpage

\begin{appendices}
\section{Extended Data}\label{secED}
\setcounter{figure}{0}
\renewcommand{\thefigure}{ED\arabic{figure}}

\begin{figure}[H]\centering
\includegraphics[width=\textwidth]{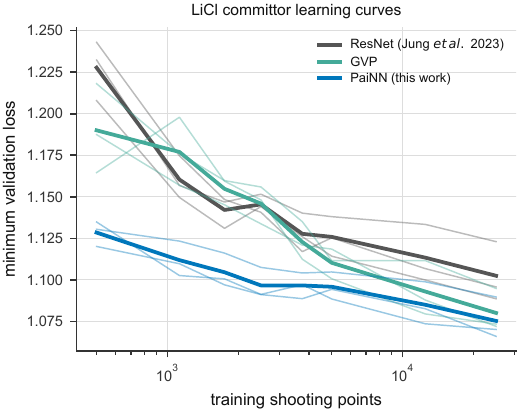}
\caption{\textbf{LiCl committor learning curves.} Minimum validation loss versus the number of training shooting points for the descriptor-free PaiNN and GVP committor models, vs the ResNet baseline of \citet{jungMachineguidedPathSampling2023} (thin lines: three replicates; thick: mean). On this ion-pair committor benchmark, the descriptor-free PaiNN committor matches or exceeds the hand-featurized baselines across training-set sizes; its calibration on the held-out validation set is shown in Fig.~\ref{fig:committor_method}d,e.}
\label{fig:ed_licl_detail}
\end{figure}


\begin{figure}[H]\centering
\includegraphics[width=0.86\textwidth]{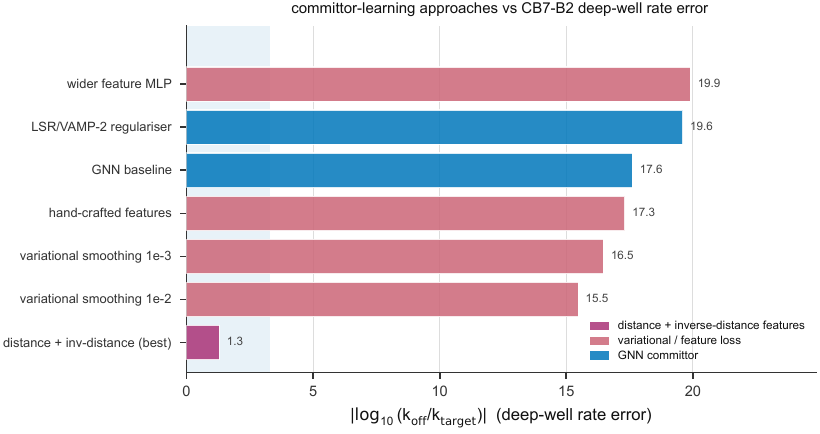}
\caption{\textbf{Negative results on committor learning (CB7-B2).} Absolute unbinding-rate error ($|\log_{10}(k_\mathrm{off}/k_\mathrm{target})|$; $k_\mathrm{target}$ = the OPES-flooding reference off-rate) for a panel of committor models from our CB7-B2 architecture/feature search, each evaluated on the full path ensemble of one AIMMD replicate. The shaded band distinguishes methods that would produce acceptable results. The models are: \emph{distance + inv-distance} — a multilayer perceptron (MLP) committor on pairwise host–guest distances and their inverses (engineered features); \emph{GNN baseline} — our descriptor-free PaiNN graph-neural-network (GNN) committor; \emph{hand-crafted features} — an MLP on a fixed hand-picked feature set; \emph{variational smoothing $10^{-2}/10^{-3}$} — the same committor with an added variational finite-lag (self-consistency) gradient penalty of that weight; \emph{wider feature MLP} — the feature MLP with a wider network; and \emph{LSR/VAMP-2 regularizer} — an auxiliary VAMP-2 slow-mode (low-rank spectral) regularizer on the committor. Only the engineered distance/inverse-distance committor reaches the reference window; all other variants — the auxiliary objectives (variational-gradient smoothing, VAMP-2 slow-mode regularization) and the wider/hand-crafted nets — leave the deep-well rate $15$–$20$ orders of magnitude off and none reliably reaches the reference window. The extra objectives scatter around the plain GNN baseline (some marginally lower, some higher in error) rather than robustly beating it, underscoring the difficulty of finding a robust ad-hoc fix for rate reweighting from committor models trained on deep-well data.}
\label{fig:ed_negative_results}
\end{figure}

\begin{figure}[H]\centering
\includegraphics[width=0.9\textwidth]{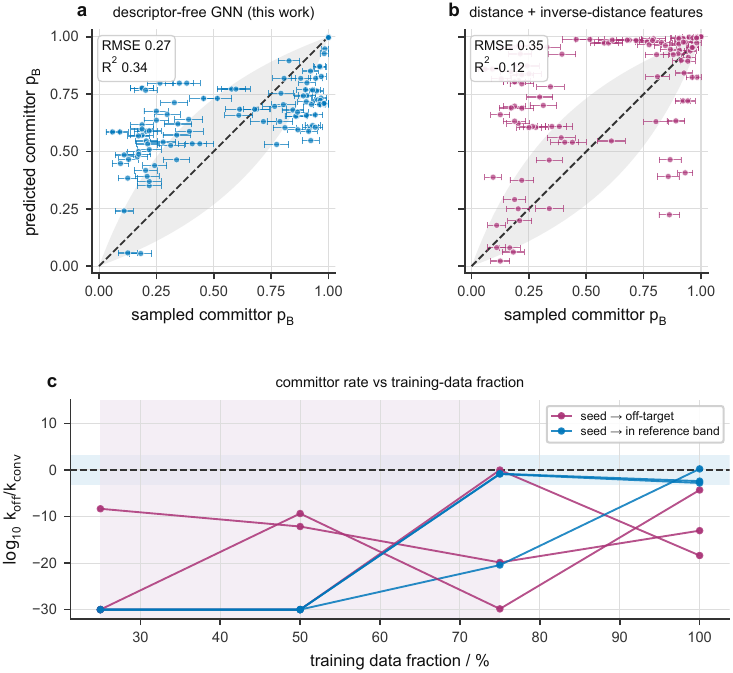}
\caption{\textbf{CB7-B2 committor-model robustness.} \textbf{a}, Descriptor-free GNN committor calibration ($R^2=0.34$). \textbf{b}, A committor built from a combination of distance-matrix and inverse-distance-matrix features achieves poorer calibration ($R^2=-0.12$); it can nonetheless occasionally recover the correct deep-well rate — but only for particular seed/sampling combinations. \textbf{c}, Committor-model off-rate deviation from the converged value versus training-data fraction; each line is one training seed, coloured by whether at full data it lands in the reference band (blue, on-target) or is off-target (purple). Rates collapse below $\sim$75\% of the reactive data and remain seed-dependent even at full data. Basin-restricted accelerated reweighting is stable across replicates instead (main text, figure \ref{fig:cb7}). Shaded band in (a) and (b) represents $\pm1$ logit.}
\label{fig:ed_cb7_committor}
\end{figure}

\begin{figure}[H]\centering
\includegraphics[width=\textwidth]{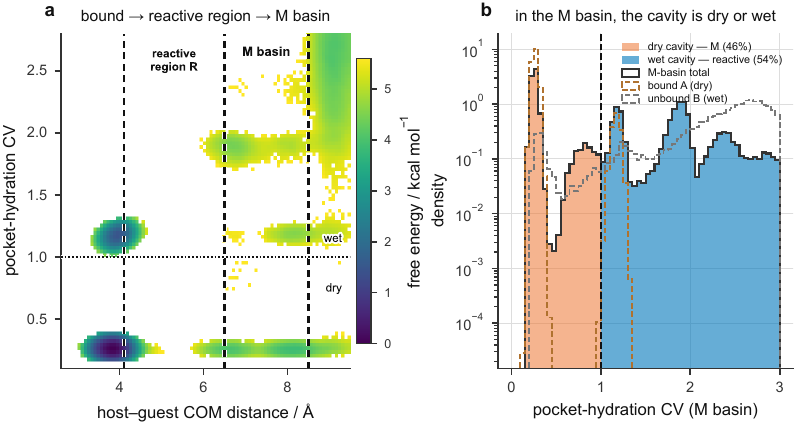}
\caption{\textbf{Calixarene bound--metastable state definitions and cavity hydration.} \textbf{a}, Unbiased-MD free-energy surface (host–guest COM distance $\times$ pocket-hydration water-density CV). The 4.1–6.5 Å band (dashed) is the reactive region R --- the barrier/transition region between the bound state A and the metastable M basin. The metastable M basin is the dry-cavity free-energy minimum at $\sim$6.5–8.5 Å: the single most-probable (lowest-free-energy) configuration there has an empty, dry cavity. \textbf{b}, Hydration distribution of frames inside the M-basin distance window (log density; the dry and wet filled histograms are normalized to the M-basin total in the black outline): at the same COM distance the cavity is dry (orange, $46\%$ — the metastable M state, guest only) or wet (blue, $54\%$ — water-filled, reactive unbinding-channel configurations), a strong bimodal degeneracy bracketed by the pure bound (dry) and unbound (wet) reference distributions (dashed outlines). A distance-only boundary cannot separate the metastable M intermediate from these reactive configurations, so the M-state definition requires a hydration (dry-cavity) term.}
\label{fig:ed_state_sensitivity}
\end{figure}

\begin{figure}[H]\centering
\includegraphics[width=\textwidth]{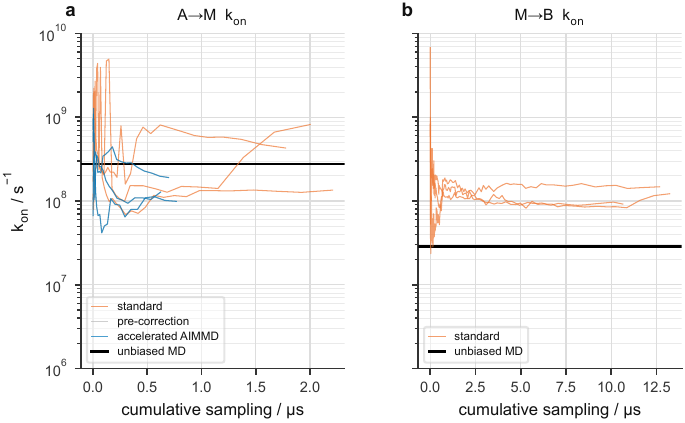}
\caption{\textbf{Calixarene-G4 partial-transition association kinetics.} On-rate ($k_\mathrm{on}$) convergence for the A→M (\textbf{a}) and M→B (\textbf{b}) sub-steps. The A→M on-rate (\textbf{a}) approaches the unbiased-MD reference (solid black), whereas for the diffuse M→B step (\textbf{b}) the standard-AIMMD on-rate levels off within a factor of $\sim$3 of the reference rather than reaching it. The corresponding A→M and M→B committor validations are shown in the main text (Fig.~\ref{fig:calixarene}e).}
\label{fig:ed_association}
\end{figure}

\begin{figure}[H]\centering
\includegraphics[width=0.95\textwidth]{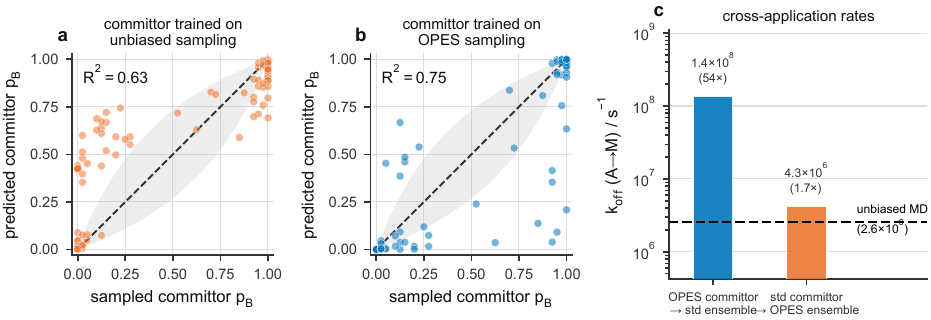}
\caption{\textbf{Committor and rate calibration: standard vs accelerated AIMMD.} \textbf{a,b}, Committor calibration for the A→M committor trained on \emph{unbiased} sampling (\textbf{a}, orange) and on \emph{OPES-biased} sampling (\textbf{b}, blue), each evaluated on the same fixed validation set: predicted committor $p_B$ (network) versus the brute-force sampled committor. The two regimes are comparably calibrated ($R^2=0.63$ unbiased vs $0.75$ OPES), so OPES-biased sampling does not by itself improve committor in the transition region. \textbf{c}, Cross-application rates: each sampling regime's committor is applied \emph{without retraining} to the \emph{other} regime's path ensemble, and compared with the unbiased-MD reference ($2.6\times10^{6}\,\mathrm{s}^{-1}$, dashed). The committor trained on OPES-biased sampling applied to the unbiased ensemble gives the overestimated $k_\mathrm{off}(\mathrm{A}\!\to\!\mathrm{M})\approx1.4\times10^{8}\,\mathrm{s}^{-1}$ (54$\times$ the reference, matching the unbiased ensemble's native estimate $1.8\times10^{8}\,\mathrm{s}^{-1}$). The committor trained on \emph{unbiased} sampling applied to the \emph{OPES-biased} ensemble (bias-reweighted) recovers $4.3\times10^{6}\,\mathrm{s}^{-1}$ (1.7$\times$ the reference) (matching the OPES run's native estimate $4.5\times10^{6}\,\mathrm{s}^{-1}$). Within each ensemble the estimate is essentially independent of which regime trained the committor: the committor is transferable across sampling ensembles. The rate is governed by the sampling ensemble and its (bias-)reweighting rather than by committor origin. Taken with the comparable calibration in panels \textbf{a,b}, this shows that a comparable committor fit near the transition region does not by itself determine the rate: OPES changes the sampling and reweighting by changing the depth up to which the committor has to be well behaved.}
\label{fig:ed_opes_control}
\end{figure}

\begin{figure}[H]\centering
\includegraphics[width=0.72\textwidth]{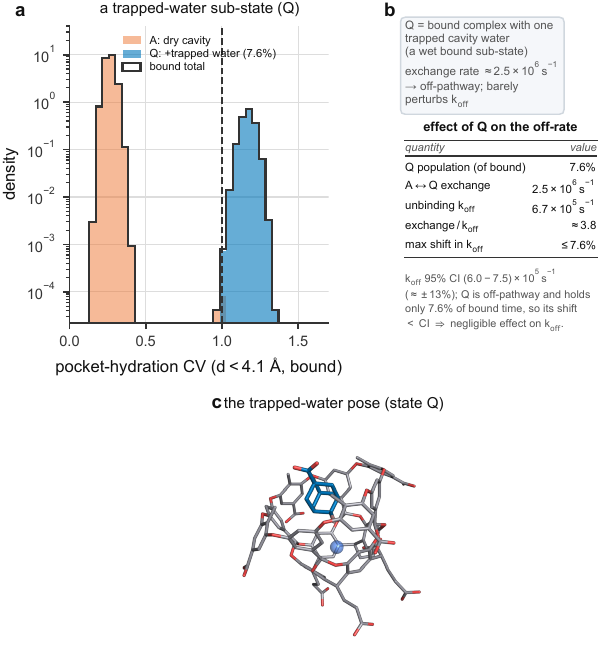}
\caption{\textbf{A trapped-water sub-state of the bound complex leaves the off-rate unchanged.} \textbf{(a)}~Distribution of the pocket-hydration (water-density) CV within the bound basin ($d<4.1$ Å). The cavity is almost always dry — only the guest is inside (state A, orange) — but in a minority of bound frames ($7.6\%$, beyond the dashed threshold) a water molecule is trapped in the cavity alongside the guest, a distinct wet-bound microstate we label \emph{Q} (blue). \textbf{(b)}~Q is \emph{off-pathway}: it sits inside the bound basin rather than on the A$\to$B unbinding coordinate, and it exchanges with the dry bound state at $\approx2.5\times10^{6}\,\mathrm{s}^{-1}$, about $3.8\times$ faster than the unbinding rate $k_\mathrm{off}\approx6.7\times10^{5}\,\mathrm{s}^{-1}$ (95\% CI $(6.0\text{--}7.5)\times10^{5}\,\mathrm{s}^{-1}$). Q therefore stays in fast equilibrium with A rather than acting as a rate-limiting intermediate. Because Q is off-pathway and carries only $7.6\%$ of the bound population, including it in the bound state can change $k_\mathrm{off}$ by at most that fraction ($\le7.6\%$) — smaller than the statistical uncertainty on $k_\mathrm{off}$ itself — so the trapped-water sub-state has a negligible effect on the measured off-rate. \textbf{(c)}~A representative Q configuration from unbiased MD: the G4 guest (blue) bound in the calixarene cavity (grey) alongside the single trapped water molecule (pale-blue sphere).}
\label{fig:ed_qstate}
\end{figure}

\begin{figure}[H]\centering
\includegraphics[width=0.92\textwidth]{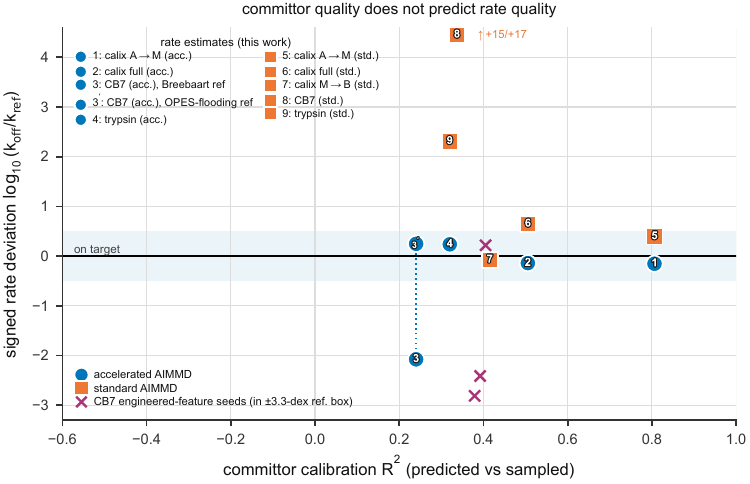}
\caption{\textbf{Committor calibration does not determine rate quality in the moderate-$R^2$ regime probed here.} Signed recovered-rate deviation ($\log_{10} k_\mathrm{off}/k_\mathrm{ref}$; $0$ = on target, positive = over-estimate) versus committor calibration ($R^2$ of predicted vs brute-force sampled committor) for every system/transition/method combination computed in this work (numbered legend). Accelerated AIMMD (blue) lands near zero across the board, whereas standard AIMMD (orange) over-shoots for most systems — dramatically for the CB7-B2 deep well (arrow, off-scale), the calixarene M$\to$B step being the sole near-target exception — at the \emph{same} committor $R^2$: within this moderate-$R^2$ regime, the method, not the committor calibration, sets the rate. Across this range, a good committor fit is neither necessary nor sufficient for an accurate rate: the trypsin committor is mediocre yet its rate lands near experiment, and the identical committor gives on- or off-target rates depending only on the method. The CB7-B2 deep-well reference is ambiguous, so the accelerated CB7 rate is shown against \emph{both} candidate references, connected by a dotted line at their shared $R^2$: point~3 uses the Breebaart/Bolhuis value ($4\times10^{-9}\,\mathrm{s^{-1}}$), while its primed sibling 3$'$ uses our independent OPES-flooding estimate ($1.9\times10^{-11}\,\mathrm{s^{-1}}$), against which the accelerated rate is on target — the two candidate references themselves differ by $\sim$2.3 dex. The off-scale standard CB7 point (arrow) over-shoots by $+15$/$+17$ dex against the Breebaart and OPES-flooding references respectively. Purple crosses are CB7 \emph{engineered-feature} committor seeds (a multi-scale soft-contact / inverse-distance MLP, \emph{not} the descriptor-free GNN) that fall inside the wide $\pm$3.3-dex deep-well reference box (the acceptance window used in Extended Data~2): at one and the same committor $R^2$ their per-seed rates scatter across that box, from on target down to $\sim$2.8 dex low, while three further seeds collapse far off-scale and are omitted — reinforcing that committor quality does not fix the rate. The shaded band marks the tighter on-target region (within $\sim$3$\times$, i.e.\ $\pm$0.5 dex, of reference).}
\label{fig:ed_committor_vs_rate}
\end{figure}
\begin{figure}[H]\centering
\includegraphics[width=0.92\textwidth]{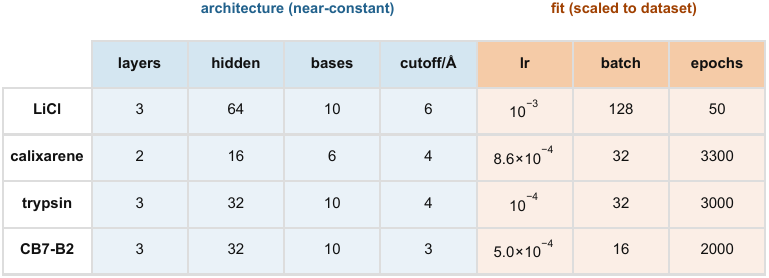}
\caption{\textbf{One descriptor-free GNN committor, minor per-system tuning.} The same PaiNN (E($n$)-equivariant message-passing) committor class is applied to all four systems, with hyperparameters read from the final production params files. The \emph{architecture} (blue: message-passing layers, hidden channels, radial bases, cutoff, dropout) is near-constant — 2–3 layers, 16–64 hidden channels, 6–10 bases, 3–6 Å cutoff, and dropout fixed at 0.1 — while only the \emph{fit} settings (orange: learning rate, batch size, epochs) are scaled to each dataset's size. No per-system input features are engineered; the graph is cached per run for efficiency. The tuning is thus minor and mechanical, not a per-system redesign.}
\label{fig:ed_architecture}
\end{figure}
\end{appendices}

\end{document}